# Influence of microstructure on mass loss caused by acoustic and hydrodynamic cavitation

J Hofmann[1,2,*], C Thiébaut[1], M Riondet[2], P Lhuissier[1], S Gaudion[3] and M Fivel[1]

[1]Univ. Grenoble Alpes, CNRS, Grenoble INP, SIMaP, F-38000 Grenoble, France
[2]Univ. Grenoble Alpes, CNRS, Grenoble INP, LEGI, F-38000 Grenoble, France
[3]Laboratory General Electric Advanced Technology, F-38000 Grenoble, France

*Corresponding author: julien.hofmann@grenoble-inp.fr

**Abstract.** The proposed study investigates the damage mechanisms of martensitic stainless steel X3CrNiMo13-4 exposed to cavitation using two complementary experimental apparatus: ultrasonic horn (MUCEF) and hydrodynamic tunnel (PREVERO). Cavitation testing has been carried out on two different metallurgical states: QT780 and QT900 corresponding to coarse and fine microstructure respectively. Acoustic cavitation erosion tests have been performed on the MUCEF equipment inspired from the ASTM G32 standards but specially designed to be installed inside X-Ray tomographs. The ultrasonic horn operates at 20 kHz and the tested specimen is located at 500 µm from the horn tip. Hydrodynamic cavitation erosion tests were conducted with classic experimental conditions of PREVERO device: a cavitation number of 0.87 corresponding to a flow velocity of 90 m.s$^{-1}$ and an upstream pressure of 40 bars. For acoustic cavitation, mass loss has been identified as dependent of the microstructure while for hydrodynamic cavitation the mass loss is identical whatever the microstructure size.

## 1. Introduction

Cavitation, *i.e.* formation of vapour bubbles in a liquid after a local pressure drop, is a very common phenomenon in hydraulic devices such as water turbines, propellers or pumps (1). These cavities develop in high-velocity regions which are usually associated to low pressures according to Bernoulli equation. Bubbles or cavitation clouds later collapse where the pressure increases. When this collapse occurs at the vicinity of a solid wall, micro-jets and/or shock waves are generated and directed toward the solid surface. Each bubble collapse is local and violent since the resulting stress can reach few gigapascals at a strain rate as high as $10^6$ s$^{-1}$ on a surface of a few squared micrometres (1). If the load generated by the collapse is greater than the yield stress, a pit is printed on the surface. Repetition of bubbles collapse increases the work-hardening of the material. This leads to damage and ultimately to mass loss (2,3). The latter depends on the type of apparatus used for testing the material with the experimental conditions as well as the type of material tested (4). The aim of the present work is to compare the effect of the microstructure of the X3CrNiMo13-4 martensitic stainless steel on the damage process for acoustic and hydrodynamic cavitation. The study proposes to compare the mass loss for the two types of microstructure for acoustic and hydrodynamic cavitation. The differences in the material response is discussed based on a microstructural analysis.

## 2. Material and methods
### 2.1. Material
In the present study, the low-carbon martensitic stainless steel X3CrNiMo13-4 was used as the tested material. The two metallurgical states QT780 and QT900, corresponding respectively to coarse and fine microstructure, have been studied. Both grades were austenitised at 1100 °C and quenched. The differences in the two grades QT780 and QT900 lays in the tempered sequence, 600 °C for 2*4 h and 540 °C for 4 h respectively. The resulting materials are composed of a $\alpha'$ martensitic matrix with lamellar reversed austenite, $\gamma_{rev}$ and residual delta ferrite, $\delta Fe$. Presence and proportion of these phases were confirmed using XRD analysis. Due to the different tempering conditions, the percentage of reversed austenite in each grade is different: respectively 11.9 vol.% and 2.7 vol.% for QT780 and QT900 grades. The reversed austenite can transform to martensite when submitted to plastic deformation: this is known as the transformation-induced plasticity (TRIP) effect. This transformation is associated to a volumetric expansion, which produces compressive forces that delay crack initiation and propagation. The low-carbon X3CrNiMo13-4 is hence widely used for water turbine manufacturing due to its high mechanical strength and high corrosion resistance. Figure 1(a) and Figure 1(b) show the orientations maps for the grades QT780 and QT900 respectively at the initial state with the martensitic matrix $\alpha'$ and reversed austenite lamellas $\gamma_{rev}$. These observations using EBSD technique reveal that the QT780 grade is coarser than the QT900 grade.

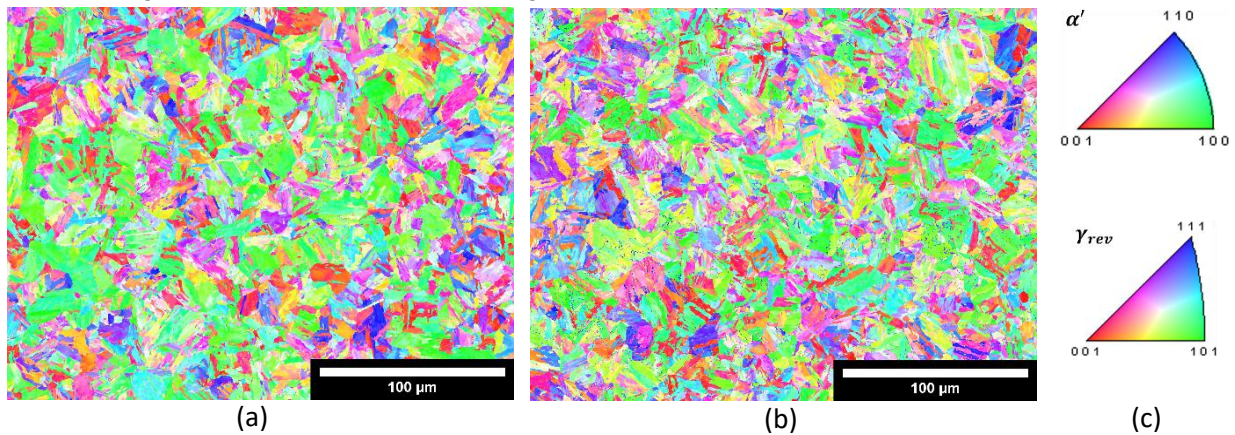

(a) (b) (c)

Figure 1: Electron backscatter diffraction (EBSD) orientation maps for grade (a) QT780 and (b) QT900 of X3CrNiMo13-4 martensitic stainless steel. (c) Inverse pole figures.

### 2.2. Cavitation erosion testing
For testing material, two experimental devices were used for generating acoustic and hydrodynamic cavitation. For MUCEF device, a sonotrode vibrating at 20 kHz with a peak-to-peak amplitude of 50 µm was used for generating acoustic cavitation. The distance of 500 µm between the specimen and tip was measured using x100 optical camera (Dino-Lite AM4815ZTL, Taiwan) and adjusted using a micrometric motorized stage (OWIS LTM120, Germany). This in-house setup, using indirect method and inspired from the ASTM G32 standard, is depicted in Figure 2(a).

Hydrodynamic cavitation testing was performed using the famous 40 bars flow tunnel called PREVERO located at LEGI laboratory. Water was accelerated using a centrifugal pump rotated by an 80 kW electric engine. Tap water was pressurized using nitrogen gas on the free surface of a water tank. As shown in Figure 2(b) on a cross section view, specimen were located at 2.5 mm from the nozzle wall where the velocity is approximately equal to 90 m.s$^{-1}$ leading to the formation of cavitation bubble clouds. Hydrodynamic cavitation erosion tests were conducted at a constant cavitation number σ = 0.870 ± 0.001.

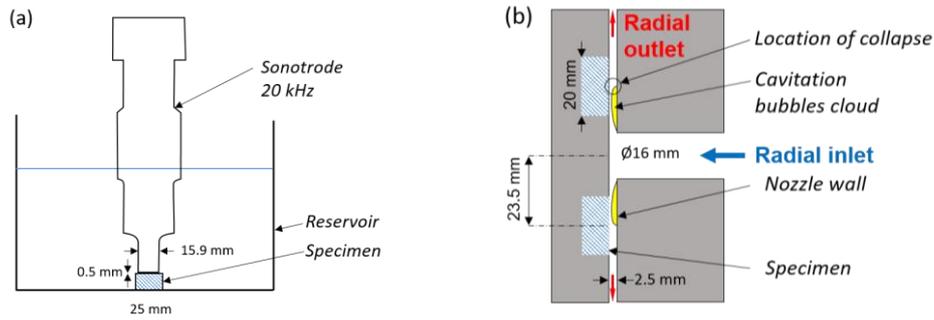

Figure 2: (a) MUCEF device used for generating acoustic cavitation. Specimen is located at 0.5 mm from the sonotrode (indirect method). (b) Cross section view of PREVERO device used for generating hydrodynamic cavitation. A constant cavitation number σ of 0.789 has been used for experiments.

## 3. Results and discussions

Figure 3 shows the mass loss as a function of the exposure time to cavitation for both grades QT780 (coarse microstructure) and QT900 (fine microstructure) for acoustic and hydrodynamic cavitation. For each measurement point, the specimen were dried carefully and then weighted using an analytical balance. Incubation time, defined as the exposure time required to observe first mass loss, is greater for hydrodynamic cavitation than acoustic cavitation. The incubation times are respectively 1 h and 2 h for the grades QT780 and QT900 for acoustic cavitation while it is approximately 5 h for both grades for hydrodynamic cavitation. For hydrodynamic cavitation, the type of microstructure has no effect on the incubation time and both mass loss curves are identical. According to MUCEF experiments, the mass loss measured for the coarser microstructure (QT780) is more important than for the finer (QT900). This is consistent with the ranking of the ultimate strengths of these two grades. It is concluded that the mass loss curves depend on the microstructure only for acoustic cavitation whereas hydrodynamic cavitation could not distinguish the behaviour of the two grades. This difference of material response according to the type of cavitation could be attributed to the size of the bubble. It has been previously shown that hydrodynamic cavitation erosion creates large pits (20-250 µm) at a small pitting rate and, conversely, acoustic cavitation erosion creates small pits (5-30 µm) at a high pitting rate (5). Bubbles size generated by acoustic cavitation is thus closer to the characteristics size of the microstructure while the collapse of hydrodynamic cavitation bubbles does not differentiate the grains.

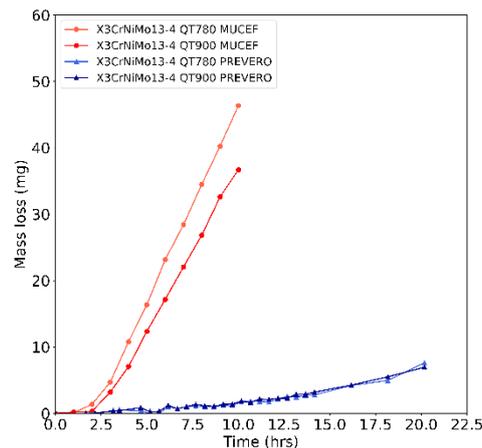

Figure 3: Mass loss as a function of exposure time in acoustic (MUCEF) and hydrodynamic cavitation (PREVERO) for the QT780 and QT900 grades.

## 4. Conclusion

Influence of the microstructure on mass loss was investigated on the X3CrNiMo13-4 martensitic stainless steel exposed to acoustic and hydrodynamic cavitation. By analysing the microstructure, we might draw the conclusion that the microstructure size has an influence on the mass loss when exposed to acoustic cavitation and not for hydrodynamic cavitation. *In-situ* synchrotron X-ray tomography analysis will be further conducted for observing cracks propagation and for contributing to the construction of a damage model.